\documentclass[apj]{emulateapj}
\usepackage{natbib}
\usepackage{amssymb}
\usepackage[displaymath]{lineno}
\usepackage{graphicx}
\usepackage{subfigure}
\usepackage{lineno}

\usepackage{amsmath}

\shorttitle{Prompt and afterglow energy in swift GRBs}
\shortauthors{Wygoda et al.}

\begin{document}

\title{The energy budget of GRBs based on updated prompt \& afterglow observations}

\author{
N.~Wygoda\altaffilmark{1,2},
D. ~Guetta\altaffilmark{3,4},
M. A. ~Mandich\altaffilmark{5}, and
E.~Waxman\altaffilmark{1}
}

\altaffiltext{1}{Department of Particle Physics \& Astrophysics,
  The Weizmann Institute of Science, Rehovot 76100, Israel}
\altaffiltext{2}{Department of Physics, NRCN, P.O. Box 9001, Beer-Sheva 84015, Israel}
\altaffiltext{3}{Osservatorio Astronomico di Roma}
\altaffiltext{4}{ORT Braude, Karmiel 21982}
\altaffiltext{5}{Department of Mathematics, Université de Bordeaux I, Talence, 33405, France}

\begin{abstract}
We compare the isotropic equivalent $15-2000$~keV $\gamma$-ray energy, $E_\gamma$, emitted by a sample of 91 \textit{swift} Gamma-Ray Bursts (GRBs) with known redshifts, with the isotropic equivalent fireball energy, $E_{\rm fb}$, as estimated within the fireball model framework from X-ray afterglow observations of these bursts. The uncertainty in $E_\gamma$, which spans the range of $\sim10^{51}$erg to $\sim10^{53.5}$~erg, is $\approx25\%$ on average, due mainly to the extrapolation from the BAT detector band to the $15-2000$~keV band. The uncertainty in $E_{\rm fb}$ is approximately a factor of 2, due mainly to the X-ray measurements' scatter. We find $E_\gamma$ and $E_{\rm fb}$ to be tightly correlated. The average(std) of $\eta^{11\rm hr}_\gamma\equiv\log_{10}(E_{\gamma}/(3\varepsilon_eE^{11\rm hr}_{\rm fb}))$ are $-0.34(0.60)$, and the upper limit on the intrinsic spread of $\eta_\gamma$ is approximately 0.5 ($\varepsilon_e$ is the fraction of shocked plasma energy carried by electrons and $E^{x\rm hr}_{\rm fb}$ is inferred from the X-ray flux at $x$ hours). If the uncertainties in the determinations of $E_\gamma$ and $E_{\rm fb}$ are twice larger than we estimated, then the data imply no intrinsic variance in $\eta_\gamma$. We also find that $E_{\rm fb}$ inferred from X-ray observations at 3 and 11 hours are similar, with an average(std) of $\log_{10}(E^{3\rm hr}_{\rm fb}/E^{11\rm hr}_{\rm fb})$ of $0.04(0.28)$. The small variance of $\eta_\gamma$ implies that burst-to-burst variations in $\varepsilon_e$ and in the efficiency of fireball energy conversion to $\gamma$-rays are small, and suggests that both are of order unity. The small variance of $\eta_\gamma$ and the similarity of $E^{3\rm hr}_{\rm fb}$ and $E^{11\rm hr}_{\rm fb}$ further imply that $\varepsilon_e$ does not vary significantly with shock Lorentz factor, and that for most bursts the modification of fireball energy during the afterglow phase, by processes such as radiative losses or extended duration energy injection, are not significant. Finally, our results imply that if fireballs are indeed jets, then the jet opening angle satisfies $\theta\geq 0.1$ for most cases. Extending our analysis to late times we find a significant reduction in $E_{\rm fb}$, $<E^{3\rm hr}_{\rm fb}/E^{2\rm d}_{\rm fb}>=1.4$, consistent with jet breaks on a 1~d time scale in a significant fraction of the bursts. These results are consistent with the main results of \citet{fw01}, which were based on a much smaller sample of GRBs.

\end{abstract}


\section{Introduction}
GRBs are the most powerful explosions in the Universe, and include the highest redshift objects observed. The widely accepted phenomenological interpretation of these cosmological sources is the so called "Fireball (FB) model" \citep{Paczynski86,Goodman86}. In "optically thin" versions of this model, the energy carried by the hadrons in a relativistic expanding wind (fireball) is dissipated through internal shocks between different parts of plasma. These shocks reconvert a substantial part of the kinetic energy to internal energy, which is then radiated as $\gamma$-rays by synchrotron and inverse-Compton radiation of shock-accelerated electrons \citep{RnM92,NPR92}. Alternatively, photospheric models for the prompt emission have been proposed \citep[e.g.][]{p+12,b13} as being significant contributors to shaping the prompt emission. Regardless of the mechanism responsible for the emission of the prompt gamma-rays, the fireball is expected to drive a shockwave into its surrounding medium, which decelerates as it encompasses an increasing amount of mass and is believed to be responsible for the afterglow emission on time scales of minutes to years \citep{pr93,mr97,w97,spn98}. Long term afterglow observations led to the conclusion \citep{rho99, sph99, r+09} that the fireball is not spherical bur rather jet-like.

The model for the afterglow phase is completely defined by the fireball energy, $E_{\rm fb}$, the jet opening angle $\theta$, the fraction of shocked plasma energy carried by electrons and magnetic field, $\varepsilon_e$ and $\varepsilon_B$, the spectral index of the power-law energy distribution of shock accelerated electrons, $p$, and the density of the plasma into which the shock expands, $n$. While afterglow observations are in general in good agreement with the model, it remains difficult to tightly constrain the parameters of the model based on observations. This is due to the fact that a complete calibration of the parameters requires a determination of the three breaks in the spectrum (cooling, peak and self absorption frequencies), which in turn requires a multi wavelength coverage of the afterglow over hundreds of days. In particular, the fireball energy, and hence the efficiency with which this energy is converted to $\gamma$-rays, can be determined in only a few cases, for which adequate spectral coverage is available \citep[e.g. GRB970503,][]{fwk00}.

\citet[][hereafter FW01]{fw01} have shown that it is possible to estimate the fireball energy (note that throughout this work, 'fireball energy', as well as GRB energy, refer to the isotropic equivalent energies) carried by electrons ($\varepsilon_eE_{\rm fb}$) using a measurement at a single time of the flux at a frequency for which the emission is dominated by fast cooling electrons (i.e. electrons for which the cooling time is shorter than the dynamical expansion time), and that this estimation is weakly dependent on poorly constrained model parameters. The physical basis underlying this estimate is that the luminosity at such a frequency corresponds to the rate at which energy is deposited in shock accelerated electrons, $\sim \varepsilon_eE_{\rm fb}/t$. Since such observations have been carried out for many GRBs, this allows one to infer the efficiency of $\gamma$-ray production for a large sample of GRBs. Using this method, it was found in FW01 that $E_\gamma$ and $\varepsilon_eE_{\rm fb}$, inferred from X-ray observations at $\sim11$ hours, are strongly correlated, implying that the burst-to-burst variations in $\varepsilon_e$ and in the efficiency of fireball energy conversion to $\gamma$-rays are small, and suggesting that both are of order unity. It was also pointed out that if GRB fireballs are indeed jet-like, then the jet opening angle should satisfy $\theta\ge0.1$ for most bursts (for smaller opening angles, sideways expansion of the jet would occur at earlier times, thus significantly affecting the inferred isotropic equivalent energy, at $t<11$ hours). While these results are significant, they were based on a small sample of GRBs and their afterglows: a total of 13 GRBs of which only 7 had measured redshifts.

The applicability of the results and conclusions of FW01 to the majority of GRBs is challenged by more recent studies, which suggest that the fireball energy is significantly modified on a time scale of hours following the burst by extended energy injection or significant radiative losses. \citet{n+06} suggested, based on an analysis of 9 lightcurves from a sample of 27 GRBs, that the fireball energy is increased by a factor $\ge4$ due to energy injection on a timescale mostly up to 3 hours (with two exception of up to 5 and 11 hours); \citet{z+06} argue that a good fraction of GRBs (though they do not quantify how many) have shallow decay phases on timescales of up to a few hours, and interpret them as corresponding to continual energy injection, which increases the fireball energy by a factor of up to 10, and \cite{pv12} find significant energy injection to be ubiquitous in afterglows through analyzing the consistency of breaks with the fireball model. Analysing two GRB afterglows \citet{bkf04} concluded, based on \citet{y+03}, that 50\% and 90\% of the fireball energy of the two bursts was radiated away during the early (3~hours and 6~days) afterglow. It was furthermore argued \citep{se01} that the X-ray flux based estimate of the fireball energy $E_{\rm fb}$ should be strongly affected by inverse-Compton (IC) energy loss of the electrons, which may lead to a significant fraction of the electron energy being radiated at (unobserved) frequencies well above the X-ray band (where the emission is dominated by synchrotron radiation), and which is expected to strongly vary between bursts.

Such significant modifications of the fireball energy (or of the estimated fireball energy, e.g. due to IC suppression of the synchrotron emission), which vary from burst to burst and are not expected to be correlated with the prompt $\gamma$-ray emission, would introduce a large scatter to the $E_\gamma/E_{\rm fb}$ ratio, which would be inconsistent with the results of FW01. In this paper we expand the analysis of FW01 to a large sample of Swift GRBs, in order to examine whether the tight correlation of $E_\gamma$ and $E_{\rm fb}$, and the implications of such a tight correlation, hold for the majority of the GRBs. The large sample and the improved afterglow data available enable us to quantify the uncertainties more reliably and thus also to draw quantitative conclusions.

The GRB sample used in our analysis is described in \S~\ref{sec:sample}. The methods used for estimating $E_\gamma$ and $E_{\rm fb}$, and the uncertainties in these estimates, are described in \S~\ref{sec:FB} and in \S~\ref{sec:GRB} respectively. The correlation between $E_\gamma$ and $E_{\rm fb}$ is analyzed in \S~\ref{sec:results}. The implications of the relatively tight correlation between $E_\gamma$ and $E_{\rm fb}$ are discussed in \S~\ref{sec:discussion}, and our conclusions are summarized in \S~\ref{sec:conclusions}.

\section{The GRB sample}\label{sec:sample}
Our sample consists of 91 long GRBs with known redshifts, which were detected by Swift in the period between May 2006 and August 2009 (from GRB060502A to GRB090812). The total number of long GRBs that were detected in this period, and for which redshifts are known, is $112$ (7 short, $T_{90}<1.8$~s, GRBs with known redshift that were detected during this period are not included in our sample). 21 of the 112 GRBs were excluded from the analysis due to the following reasons: for 1 GRB the required BAT data are not available; 14 GRBs do not have satisfactory XRT data at the 3-11 hours time interval (For 2 of these bursts no XRT data are available, for 8 bursts no data are available at $t>4$~hr, and for 4 there are only 3 data points in the time range of $3{\rm hr}<t<11$~hr. We note that the lack of data for these bursts is unlikely to be due to low X-ray fluxes, and is likely due to low sampling cadence, since the existing data points show relatively high fluxes.); 6 bursts have late time (close to 3 hours) flares or clearly rising lightcurves around the relevant times.

The exclusion of the latter group of 6 GRBs from our analysis implies that our conclusions may not apply to a small minority, $\sim5\%$, of the long GRB population.

\section{Estimating $E_{\rm fb}$}\label{sec:FB}

The fireball energy carried by electrons can be estimated from a measurement of the flux density $f_{\nu}$ at a frequency $\nu$, which is above the cooling frequency (corresponding to the frequency of radiation emitted by electrons with cooling time equal to the dynamical time), using equations 4 and 5 of FW01:
\begin{equation}
\label{eq:EFB}
\varepsilon_eE_{\rm fb}=(C_2 C_3)^{-1/2}C_1^{-1}\frac{d_L^2}{1+z}\nu t f_{\nu}(\nu,t){Y^{\varepsilon}},
\end{equation}
where
\begin{equation}
\label{eq:C13}
Y\equiv C_1 C_3^{1/2}C_2^{-3/2}\varepsilon_e^{-3}\varepsilon_B^{-1}d_L^{-2}\nu t^2 f_{\nu}^{-1}(\nu,t), \,\,\,\,\,\,
\varepsilon\equiv\frac{p-2}{p+2}\,
\end{equation}
and $C_1,C_2,C_3$ are numerical constants taken from FW01, $C_1=1.4\times 10^{-21} {\rm cm}^{3/2}$, $C_2=6.1\times 10^{-5} {\rm s^{3/2} g^{-1/2} cm^{-1}}$ (for $p=2.2$) and $C_3=6.9\times 10^{39} {\rm s^{-3/2}g^{-1/2}{\rm cm}^{-2}}$.
The inferred energy is independent of the density of the plasma into which the fireball expands, and very weakly dependent on the value of $\varepsilon_B$ for $p\approx2$. For our nominal energy estimates we use $\varepsilon_e= \varepsilon_B=1/3$. The dependence on $\varepsilon_e$ is shown explicitly in all our results.

For the spectral index, we adopt in our analysis a value of $p=2.2$. We show below that such a universal value is supported by both the afterglow spectra (\S~\ref{sec:p}) and the time dependence of afterglow X-ray flux (\S~\ref{sec:Xray_flux}). As discussed in some detail in \S~\ref{sec:micro}, the tight correlation between $E_{\rm fb}$ and $E_\gamma$ further supports a universal value of $p$. Using a value of $p=2.4$ instead of $p=2.2$ increases the estimate of $E_{\rm fb}$ by 60\% (FW01).

Eqs.~(\ref{eq:EFB}-\ref{eq:C13}) yield an accurate estimate of $\varepsilon_e E_{\rm fb}$ under the assumptions that the emitting plasma is well described by the self-similar spherical fireball model and that electron acceleration is well characterized by time independent $\varepsilon_e$ and $p$. When these assumptions hold, the X-ray flux should drop with time as $f_X\propto t^{(2-3p)/4}$, and $E_{\rm fb}$ inferred using eq.~(\ref{eq:EFB}) should be independent of $t$. A measure of the accuracy of the determination of $\varepsilon_e E_{\rm fb}$, and a test of the validity of the above assumptions, is therefore obtained by comparing the values of $\varepsilon_e E_{\rm fb}$ inferred from the X-ray flux at significantly different times. When the assumptions described above are not valid, eqs.~(\ref{eq:EFB}-\ref{eq:C13}) provide a less accurate estimate of $\varepsilon_e E_{\rm fb}$.

We show in \S~\ref{sec:Xray_flux} that the values of $E_{\rm fb}$ inferred using eq.~(\ref{eq:EFB}) at 3~hr and 11~hr are similar (implying $d\log f_X/d\log t\simeq-1$): we find that $<\log_{10}(E^{3\rm hr}_{\rm fb}/E^{11\rm hr}_{\rm fb})>=0.04$ and that only $\approx20\%$ of the bursts have $E^{3\rm hr}_{\rm fb}$ and $E^{11\rm hr}_{\rm fb}$ values differing by more than a factor of 2. These results support the validity of the model assumptions described above. They may appear to be contradictory to the claims that "flat" X-ray lightcurves (with $d\log f_X/d\log t$ significantly larger than -1), and hence significant energy injection, are common \citep[e.g.][]{n+06}. This issue is discussed in \S~\ref{sec:flat}.

Finally , we note that the uncertainty in the energy estimate should include a systematic uncertainty due to the uncertainty in the values of the constants $\{C_1, C_2, C_3\}$, which vary by factors of a few between different models of the afterglow synchrotron emission \citep[see for example the discussion in][]{gkp06}. We do not include in our analysis a treatment of this systematic uncertainty, since a modification of $\{C_1, C_2, C_3\}$ will modify the values of $E_{\rm fb}$ and of $E_\gamma/E_{\rm fb}$ by a fixed factor for all bursts, but will not affect the fractional scatter of these values. It will thus not affect the main conclusions of this work.




\subsection{Constraints on $p$ from afterglow spectra}\label{sec:p}

Following FW01, we present below evidence for the validity of the approximation of $p\approx2$ based on the spectral index of the afterglow emission, determined from the X-ray and optical fluxes, $\beta_{ox}\equiv -\ln(f_x/f_o)/\ln(\nu_x/\nu_o)$. At times when the cooling frequency is below the X-ray band, we expect $(p-1)/2<\beta_{ox}<p/2$ (with the lower limit obtained when the cooling frequency is in the X-ray band, and the upper limit when it is in the optical band).

\begin{figure}[h]
\epsscale{1} \plotone{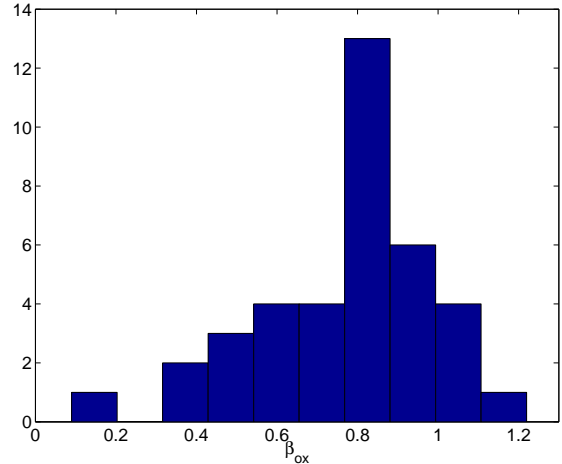}
\caption{A histogram of the effective spectral index $\beta_{ox}$ for a sub-set of 38 bursts within our sample, given by \cite{zdw09}.
\label{fig:betaox}}
\end{figure}
In fig.~\ref{fig:betaox} we show $\beta_{ox}$ at 11 hours for the first 38 GRBs in the sample (up to GRB071031), as given by \cite{zdw09}. For $75\%$ of the bursts, $\beta_{ox}$ is in the range of 0.6-1.1, which is the range expected for $p=2.2$. Of these bursts for which the measured spectral index does not lie in this range, one has $\beta_{ox}>1.1$ while the others have $\beta_{ox}<0.6$. Such values are in principle not consistent with the shape of the spectrum in the fireball model for $p=2.2$ at times when the cooling frequency lies below the measured X-ray band. However, as already noted by FW01, dust extinction in the host galaxy, which suppresses the optical flux, can reduce the observed value of $\beta_{ox}$. \cite{zdw09} find that low values of $\beta_{ox}$ typically correspond to high values of hydrogen column density, which serves as a proxy for the strength of local dust extinction. Moreover, they state (referring to previous work) that for several of these low $\beta_{ox}$ bursts, detailed studies of the optical afterglow allowed one to measure the optical extinction values, thus raising the value of $\beta_{ox}$. Finally, we have compared the group of bursts with low $\beta_{ox}$ in the sample to the whole sample and found both groups to have the same average $E_{\rm fb}$ and $E_\gamma/E_{\rm fb}$, thus indicating that the the low $\beta_{ox}$ bursts do not represent a separate population.

\subsection{$f_X$ and $E_{\rm fb}$ at 11 and 3 hrs}\label{sec:Xray_flux}

For each GRB we have estimated the 10 keV flux density at 11 and 3 hours after the GRB trigger by fitting a power-law temporal behavior to the Swift X-ray lightcurve between $3\times10^3$~s and $6\times10^4$~s. Due to the rapid decay and to the flares often characterizing the X-ray light curve at early time, we do not use $t<3\times10^3$~s data (except for 4 GRBs, which clearly show a regular behavior of the lightcurves at earlier times and for which the earlier time data are necessary in order to obtain a reliable interpolation at 3 hours). Due to the possible appearance of a jet-break suppression of the flux at late time, $t>1$~d, we do not use $t>6\times10^4$~s data (except for 12 GRBs, for which a late break in the lightcurve is clearly absent and for which later data are necessary to obtain a reliable interpolation at 11 hours).

\begin{figure}
\epsscale{1} \plotone{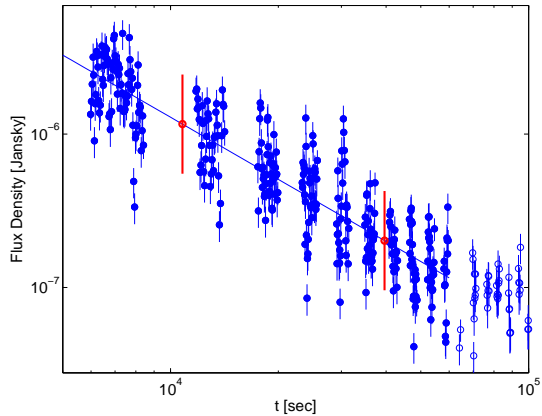}
	\caption{XRT flux measurements for GRB090618. The straight line is the power-law time decay fit for the range of interest, based on which the fluxes at 3 and at 11 hours are inferred (red circles). The red lines represent the estimated flux uncertainties (see text). The empty circles represent data points at $t>6\times10^4$~s, which are not taken into account in determining the fluxes at 3 and 11 hours.}
	\label{fig:xrt_090618}
\end{figure}

The uncertainties in the derived X-ray fluxes at 3 and 11 hours are dominated by the scatter in the reported X-ray flux measurements, as illustrated in figure~\ref{fig:xrt_090618}. We use this scatter as indicative of the flux uncertainty, rather than the reported flux uncertainties, since the latter do not appear to be consistent with the scatter of the data points. The fluxes at 3 and 11 hours are estimated using a least square fitting of a first degree polynomial in the $\log t-\log f_X$ plane, and their uncertainties are determined such that $90\%$ of the data points lie within the uncertainty. We note that for $20\%$ of the bursts, for which the last measured X-ray data point is at a time earlier than 5 hours after the trigger, the extrapolation of the power law fit to 11 hours makes a significant contribution to the uncertainty.

\begin{figure}
\epsscale{1} \plotone{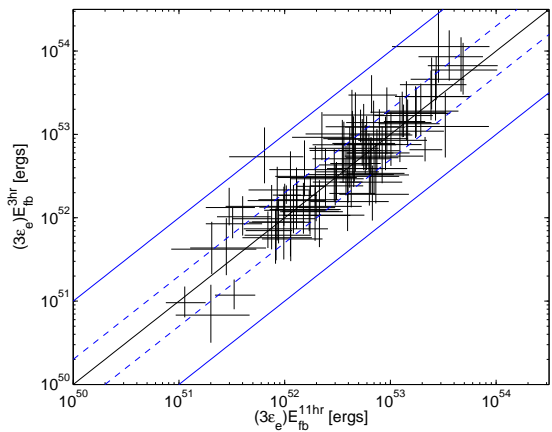}
	\caption{The relation between $E_{\rm fb}$ inferred from the X-ray fluxes at 3 and 11 hrs. The dashed (solid) blue lines correspond to the energy estimates at 3 and 11 hrs being within a factor of 2 (10) of each other.}
	\label{fig:EFB113}
\end{figure}

The average(std) of $d\log f_X/d\log t$ we obtain from our fits to the X-ray light curves between $3\times10^3$~s and $6\times10^4$~s are -1.25 (0.52), consistent with the value of -1.1 expected for $p=2.2$. The deviation of the decay rate from this value is larger than the estimated error in the determination of the decay rate for $<20\%$ of the bursts. This result is illustrated in figure~\ref{fig:EFB113}, showing the tight correlation between the fireball energies estimated at 3 and 11 hours. The average(std) of $\log_{10}(E^{3\rm hr}_{\rm fb}/E^{11\rm hr}_{\rm fb})$ are $0.04(0.28)$. Given our estimated uncertainties in inferring $E_{\rm fb}$, the data are consistent with no intrinsic variance in the ratio $E^{3\rm hr}_{\rm fb}/E^{11\rm hr}_{\rm fb}$. Out of the 91 bursts in the sample, 11(14) have a fireball energy estimated at 3 hours which is less than half (more than twice) that estimated at 11 hours. We show in \S~\ref{sec:results} that the average of $\eta_\gamma$ is similar for bursts with small (less than a factor of 2) and large (more than a factor of 2) differences between $E^{3\rm hr}_{\rm fb}$ and $E^{11\rm hr}_{\rm fb}$, while the variance of $\eta_\gamma$ is somewhat larger for the latter group.

\subsection{Flat light curves}\label{sec:flat}

The results reported in the preceding paragraph may appear to be contradictory to the claims that "flat", i.e. $d\log f_X/d\log t$ significantly larger than -1, X-ray lightcurves (and hence significant energy injection) are common \citep[e.g.][]{n+06}. This is, however, not (necessarily) the case. \citet{n+06} concluded that the canonical lightcurve contains a shallow decay phase based on an analysis of 27 bursts out of which only 9 showed flat decay phases, and only 5 ($<20\%$) showed such behavior extending to $t>6\times10^3$~s. Our analysis, based on a much larger and complete sample, shows that the fraction of afterglows showing a flat phase leading to a factor 2 or larger increase in the inferred $E_{\rm fb}$ between 3 and 11~hr is $\lesssim10\%$.

At times earlier than $3\times10^3$~s, the lightucrves are in general less regular, as was discussed in \S~\ref{sec:Xray_flux}, with many of the lightcurves showing both steeper and flatter time dependence compared to later times. In particular, roughly $25\%$ of the bursts show lightcurves which are significantly flatter at early times compared to late times. However, as we explain in detail in \S~\ref{sec:injection}, the contribution of possible early energy injection to the total fireball energy cannot in general be significant.



\section{Estimating $E_\gamma$}\label{sec:GRB}
The $\gamma$-ray spectrum of GRBs is typcially described using a Band function~\citep{b+93},
\begin{equation}
\frac{dN}{dE} = \begin{cases} A (\frac{E}{50 KeV})^{\alpha} e ^ {-\frac{E(2+\alpha)}{E_{\rm peak}}} &\mbox{if}  E < E_{\rm break} \\
A (\frac{E_{\rm break}}{50 keV\cdot e})^{(\alpha-\beta)} (\frac{E}{50 keV})^{\beta} &\mbox{if}  E \geq E_{\rm break} \end{cases}
\end{equation}\
where $E_{\rm break} \equiv \frac{(\alpha - \beta)E_{\rm peak}}{(2+\alpha)}$. Due to the limited energy range of the BAT detector, 15 to 150 keV \citep{b+05}, the typical values inferred for the Band function parameters based on BAT data differ from the typical values inferred based on data from BATSE, which is sensitive in the range of 30 to 2000~keV \citep{k+06}. While the average low energy index $\alpha$ in the BAT sample used here, $-1.11$, is similar to the average value of the BATSE GRBs, $-1.08$ \citep{k+06}, the peak energy $E_{\rm peak}$ and the high energy index $\beta$ inferred from the BAT and from the BATSE data are quite different. In particular, the average peak energy obtained for BATSE spectra is $260$~keV \citep{k+06}, outside the BAT energy band. Thus, in order to determine $E_\gamma$ in the energy range of 15 to 2000~keV for the BAT GRBs, we use a Band function with $\alpha$ and flux normalization as given in the Swift data archive (determined by the BAT spectrum), but use the average values of the BATSE catalog, $E_{\rm peak}=260$~keV and $\beta=-2.4$ \citep{k+06}.
This procedure introduces of course an uncertainty in the estimated value of $E_\gamma$, due to the uncertainty in the extrapolation of the spectrum up to 2000~keV. We estimate the uncertainties using extrapolations to high energy with the extreme values obtained in the BATSE catalog for $E_{\rm peak}$ and $\beta$: 150~keV and 500~keV for $E_{\rm peak}$ and $-2.6$ and $-2.0$ for $\beta$.

The method for estimating $E_\gamma$ and its uncertainty depends on the values of the Band function parameters given in the Swift data archive, as explained in detail below. We note that in all cases, the low end of the model spectrum we adopt is identical to that found by BAT (and its flux integrated over the 15-150~keV range matches the BAT reported integrated flux).
\begin{itemize}
\item For $\beta$ in the range of [-4,-2], which holds for $\approx15\%$ of the bursts in our sample, the flux per logarithmic photon energy interval, $E^2dN/dE$, is decreasing with $E$. We therefore estimate the [15,2000]~keV fluence by integrating the BAT Band function up to 2000~keV, and the uncertainty range by extrapolating the BAT spectrum beyond 150~keV using a $\beta=-2.6$ and a $\beta=-2.0$ power-laws.
\item For $\beta>-2$, which holds for $\approx55\%$ of the bursts, $E^2dN/dE$ is increasing with $E$. We therefore estimate the [15,2000]~keV fluence by extrapolating the BAT Band function to 260~keV, followed by a $\beta=-2.4$ power-law at higher energy. The lowest value of energy uncertainty range is obtained by extrapolating the BAT spectrum beyond 150~keV using a $\beta=-2.6$ power-law, and the highest value by extrapolating the BAT Band function to 500 keV followed by a $\beta=-2.0$ power-law at higher energy.
\item For $\beta<-4$, which holds for $\approx30\%$ of the bursts, $E_{\rm break}$ inferred by the SWIFT analysis is well above the BAT band. We therefore estimate the [15,2000]~keV fluence by extrapolating the BAT Band function to 260~keV, followed by a $\beta=-2.4$ power-law at higher energy. The lowest value of the energy uncertainty range is obtained by extrapolating the BAT Band function to 500 keV followed by a $\beta=-2.6$ power-law, and the highest value by extrapolating the BAT spectrum beyond 150~keV using a $\beta=-2.0$ power-law.
\end{itemize}

The ratios of the [15,2000]~keV fluences, inferred as described above, and the BAT [15 150]~keV fluences are in the range of 1.02 to 4.26, with an average of $2.25$. The uncertainties due to the extrapolation are $\approx25\%$ on average, reaching a factor of $2$ for some bursts. In addition to the uncertainty introduced by the extrapolation to high energies, we include in our analysis the uncertainty in the BAT fluence as reported in the SWIFT archive. The average value of the latter uncertainty is 7\%, and it amounts to no more than 26\% for any burst in the sample.

\section{The correlation between $E_\gamma$ and $E_{\rm fb}$}\label{sec:results}
\begin{figure}[h]
\epsscale{1} \plotone{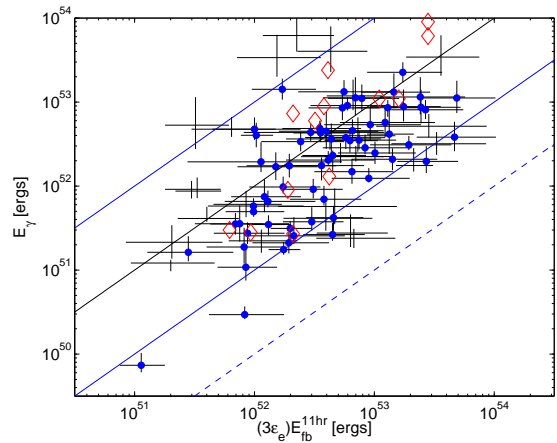}
\caption{The relation between $E_\gamma$ and $E^{11\rm hr}_{\rm fb}$. The GRBs marked with filled circles are those for which the fireball energy estimates at 3 hours and 11 hours lie within a factor 2 of each other. We show for comparison the results obtained by FW01 (red diamonds). Solid lines correspond to $E_{\gamma}/(3\varepsilon_e)E^{11\rm hr}_{\rm fb}$=0.1,1,10 and the dashed line to $E_{\gamma}/E(3\varepsilon_e)E^{11\rm hr}_{\rm fb}=0.01$.
\label{fig:EFB11hr}}
\end{figure}

Using the methods described in the preceding sub-sections we find the average(std) of $\log_{10}(E_{\gamma}/1{\rm erg})$ and of $\log_{10}(3\varepsilon_eE^{11\rm hr}_{\rm fb}/{\rm 1 erg})$ to be $52.3(0.7)$ and $52.5(0.6)$ respectively, very close to the values inferred in FW01. As can be seen in figure~\ref{fig:EFB11hr}, $\log(E_{\gamma})$ and $\log(E_{\rm fb})$ are linearly correlated,
with a high correlation coefficient of $\approx 0.6$. For the average(std) of $\eta^{11\rm hr}_\gamma\equiv\log_{10}(E_{\gamma}/(3\varepsilon_eE^{11\rm hr}_{\rm fb}))$ we find $-0.34(0.60)$, consistent with the results of FW01, $0.01(0.5)$ \citep[and also with those of][who give their nominal results with a normalization of $\varepsilon_e=0.1$]{d+12}.

Given the estimated uncertainties in the derived values of $E_\gamma$ and of $\varepsilon_eE^{11\rm hr}_{\rm fb}$, the hypothesis that their ratio is universal (i.e. the hypothesis of no intrinsic variance in $\eta^{11\rm hr}_\gamma$) is inconsistent with the data (resulting in $\chi^2$ per degree of freedom of $\approx4$). In order to estimate the intrinsic variance of $\eta^{11\rm hr}_\gamma$, we assume that it follows a gaussian distribution with a variance $s$. $s$ is determined by equating $\chi^2$ to 1 in the equation $\chi^2=(N-1)^{-1}\Sigma\frac{y_i-x_i-\eta_\gamma}{\sigma_i^2+s^2}$, where $x_i, y_i$ are $\log(E_{\gamma}), \log(E_{\rm fb})$, and $\sigma_i$ represents the uncertainty in the energy estimates ($\sigma_i^2=\sigma_{x_i}^2+\sigma_{y_i}^2$) (minimizing $\chi^2$ while setting $s=0$ gives the nominal value of $\eta_{\gamma}$, which is then used in computing $s$). We find the intrinsic variance to be approximately $0.5$. We note, though, that this value is sensitive to the uncertainties in $E_\gamma$ and $E_{\rm fb}$. For example, if these uncertainties are twice larger than we estimated, then the data implies no intrinsic variance in $\eta_\gamma$.

\begin{table}
\caption{\bf Average values and Standard deviations}

\begin{tabular}{|l|cccc|}
\hline
time                                        & 3 hr  & 3 hr  & 11 hr & 11 hr \\
set~\footnote{"all"= the entire burst sample analyzed in this work, "tight"= a subset for which the fireball energy estimates at 3 hours and at 11 hours are within a factor of 2 of each other.}
                                            & all   & tight & all   & tight \\
sample size                                 & 91    & 66    & 91    & 66    \\
\hline
$<\log_{10}((3\varepsilon_e)E_{\rm fb})>$        & 52.6  & 52.6  & 52.5  & 52.6  \\
var(${\log_{10}((3\varepsilon_e)E_{\rm fb})}$) & 0.6   & 0.6   & 0.6   & 0.5   \\
$<\eta_{\gamma}>$                  & -0.36 & -0.43 & -0.34 & -0.41 \\
var($\eta_{\gamma}$)          & 0.55  & 0.51  & 0.60  & 0.51  \\
$\chi^2$ pdf~\footnote{$\chi^2$, as defined in the text.}          & 3.3   & 2.9   & 3.6   & 2.6   \\
intrinsic variance~\footnote{required to obtain $\chi^2=1$.}       & 0.41  & 0.36  & 0.50  & 0.36  \\
\hline
\end{tabular}
\label{table:efficiencies}
\end{table}

Repeating the above analysis using X-ray observations at 3 hours, we find the average(std) of $\log_{10}(3\varepsilon_eE^{3\rm hr}_{\rm fb}/{1\rm erg})$ and $\eta_\gamma^{3\rm hr}\equiv\log_{10}(E_\gamma/3\varepsilon_eE^{3\rm hr}_{\rm fb})$ to be $52.6(0.6)$ and $-0.36(0.55)$ respectively, similar to the values estimated at 11 hours.

\begin{figure}[h]
\epsscale{1} \plotone{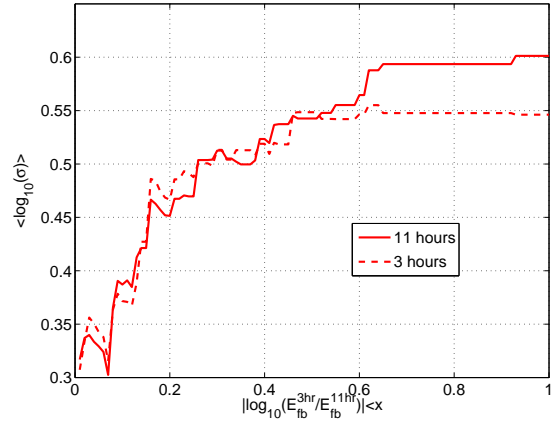}
  \caption{The standard deviation of $\eta_\gamma$ of that part of the bursts within the sample for which the fireball energy estimated at 3 and 11 hours is within a factor of $10^x$.}\label{fig:sigma_vs_tightness}
\end{figure}

\begin{figure}[h]
\epsscale{1} \plotone{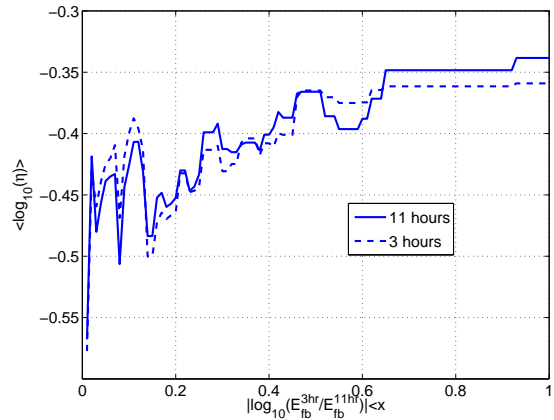}
	\caption{The average value of $\eta_\gamma$ of that part of the bursts within the sample for which the fireball energy estimated at 3~hrs and at 11~hrs are within a factor of $10^x$ of each other.}
	\label{fig:eta_vs_tightness}
\end{figure}

Table~\ref{table:efficiencies} gives the values of the average and variance of the fireball energy and of $\eta_\gamma$, obtained with and without bursts for which the fireball energy estimates at 3 and 11 hours differ by more than a factor of 2. As can be seen from the table, the results obtained with and without these bursts are very similar. We note that, as shown in Fig.~\ref{fig:sigma_vs_tightness}, the variance of $\eta_\gamma$ is larger for bursts characterized by larger differences between $E^{3\rm hr}_{\rm fb}$ and $E^{11\rm hr}_{\rm fb}$ (while its average value is nearly independent of this difference, Fig.~\ref{fig:eta_vs_tightness}).

\begin{figure}[h]
\epsscale{1} \plotone{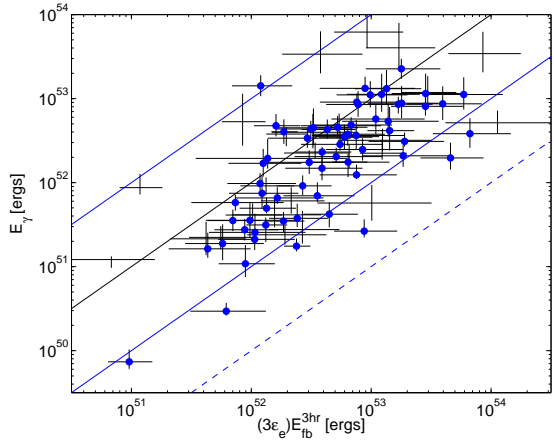}
	\caption{Same as fig.~\ref{fig:EFB11hr}, for $E_{\rm fb}$ inferred from the X-ray flux at 3~hrs (instead of at 11~hrs).}
	\label{fig:EFB3hr}
\end{figure}

\begin{figure}[h]
\epsscale{1} \plotone{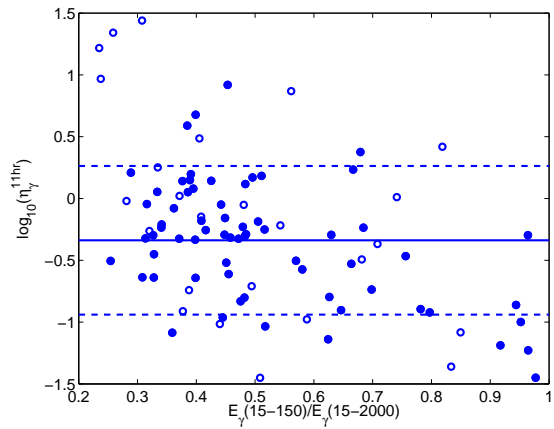}
	\caption{$\eta_\gamma$ versus the fraction of the total estimated GRB energy which is contained in the BAT range. Empty circles represent bursts for which $E^{11\rm hr}_{\rm fb}$ and $E^{3\rm hr}_{\rm fb}$ differ by more than a factor of 2. The solid line shows the average of $\eta_\gamma$, the dashed lines are separated from the average by the variance of $\eta_\gamma$.}
	\label{fig:eta_vs_GRBmeasuredfrac}
\end{figure}

While $\eta_\gamma$ lies within the range of $\sim0.1$ to $\sim1$ for most of the GRBs in our the sample, there are 5 GRBs (GRB080607, GRB080319B, GRB061110B, GRB061007, GRB060510B) with exceptionally large values of $\eta_\gamma$, $\eta_{\gamma}(3\varepsilon_e)^{-1}>8$. Out of these, $4$ bursts have exceptionally steep lightcurves between 3 and 11 hours relative to what is expected from the fireball model. One possible explanation for such behavior could be that these GRBs suffer from significant radiative losses at these times, implying that their $E_\gamma/E_{\rm fb}$ is actually lower than inferred by using the estimates of $E^{3\rm hr}_{\rm fb}$ and $E^{11\rm hr}_{\rm fb}$
(Estimating the radiative losses for particular bursts must be done using extensive lightcurve fitting to constrain the fireball model parameters, which is beyond the scope of this work).
It should however be pointed out that, as shown in Fig.~\ref{fig:eta_vs_GRBmeasuredfrac}, for these bursts the ratio of our inferred [15,2000]~keV fluence to the [15,150]~keV measured BAT fluence is 4, the largest in our sample. This implies that the $E_\gamma$  and $\eta_\gamma$ estimates of these bursts are more sensitive to systematic errors in our extrapolation method.

\section{Discussion}\label{sec:discussion}

As mentioned in the introduction, significant modifications of the fireball energy, or of the estimated fireball energy, which vary from burst to burst and are not expected to be correlated with the prompt $\gamma$-ray emission, would introduce a large scatter to the $E_\gamma/E_{\rm fb}$ ratio. Thus, the relatively tight correlation we find (\S~\ref{sec:results}) between $E_\gamma$ and $E_{\rm fb}$ sets relatively tight constraints on the combined effect of processes that lead to such modifications. Quantitatively, the intrinsic variance in $\eta_\gamma$, which is inferred from the scatter in the $E_\gamma$ - $E_{\rm fb}$ correlation, sets an upper limit of a factor of 2-3 on the scatter of the $E_\gamma/E_{\rm fb}$ ratio due to the combined effects of these processes. In what follows we discuss the main implications of the tight correlations we find between $E_\gamma$ and $E_{\rm fb}$ and between $E^{11\rm hr}_{\rm fb}$ and $E^{3\rm hr}_{\rm fb}$.

\subsection{Jet opening angle and "off-axis" detections}\label{sec:angle}
The fact that the fireball energy estimates at 3 and 11 hours are similar indicates that in general jet breaks do not occur well before 11 hours, since that would cause the inferred $E^{11\rm hr}_{\rm fb}$ to be on average significantly lower than $E^{3\rm hr}_{\rm fb}$. Since the relation between the jet opening angle $\theta_j$ and the jet break time $t_j$ is estimated as \citep[e.g.][]{lw00}
\begin{equation}
    \theta_j\approx 0.12\left(E_{\rm iso,53}/n_0\right)^{-1/8}t_{j, day}^{3/8}\,,
\label{eq:theta_lw} \end{equation}
we conclude that typical opening angles are greater than 0.1, as inferred in a similar manner by FW01.

A similar argument also implies that the population of observed GRBs can not be dominated by jets observed initially (during $\gamma$-ray emission) "off-axis", i.e. with an angle $\theta$ between the line of sight and the jet axis, which is larger than $\theta_j$. In such cases, the evolution of the fireball and of the emission of the afterglow radiation are not described by the simple self-similar model leading to eqs.~(\ref{eq:EFB}-\ref{eq:C13}), which are used to infer $E_{\rm fb}$, the luminosity of the prompt $\gamma$-ray emission would vary strongly depending on $\theta$~\citep[e.g.][]{lpr13}, and the tight correlation between $E_\gamma$ and $E_{\rm fb}$ would be a coincidence.

In order to test for the possible presence of jet breaks at $t>11$~hr, and to demonstrate the sensitivity of our analysis method to such breaks, we have derived $E_{\rm fb}^{2\rm d}$ for the 73 bursts within the sample for which X-ray light curves extending to $t>2$~d (with at least 2 measurement points at $t>1$~d) are available, using data points at $t>1$~d only. For these bursts we find $<\log_{10}(E_{\rm fb}^{3\rm hr}/E_{\rm fb}^{2\rm d})>=0.16$, significantly larger than the uncertainty in estimating $<\log_{10}E_{\rm fb}>$, which is $\approx0.07$. The hypothesis that the average energies at $3$~hr and $2$~d are equal is rejected at a $\approx85\%$ confidence level based on a standard two sample t-test (Recall that in \S~\ref{sec:FB} we have found that $<\log_{10}(E_{\rm fb}^{3\rm hr}/E_{\rm fb}^{2\rm d})>=0.04$, consistent with no change in $<\log_{10}E_{\rm fb}>$ between 3~hr and 11~hr). A possible explanation of the $\approx40\%$ reduction in $<E_{\rm fb}>$ betweenn 3~hr and and 2~d is a jet break at $t<2$~d. A precise characterization of the nature of jet breaks is beyond the scope of this work, but the $\approx40\%$ drop could correspond (in two extreme cases) to a universal jet break time of $\approx30$~hr or to $\approx1/2$ the bursts having a jet break at 11~hr.

\subsection{Variations of the $\gamma$-ray emission on small angular scales}\label{sec:dtheta}
Due to the relativistic expansion of the fireball, the flux we observe is obtained from a conical section of the fireball (around the line of sight) with an opening angle of $1/\Gamma$, where $\Gamma$ is the (time dependent) expansion Lorentz factor. The GRB $\gamma$-rays are expected to be emitted at the highly relativistic, $\Gamma=\Gamma_\gamma\sim10^{2.5}$, phase of fireball expansion, while the X-rays observed at $\sim10$~hrs are expected to be emitted after deceleration to $\Gamma=\Gamma_X\sim10^1$. Thus, the isotropic equivalent fireball energy, $E_{\rm fb}$, is determined by an average of the jet properties over angles $\sim1/\Gamma_X\sim0.1$, much larger than those from which $\gamma$-ray emission is observed, $\sim1/\Gamma_\gamma<0.01$. If the emission of $\gamma$-rays were to vary significantly over angular scales $0.01\sim1/\Gamma_\gamma<\Delta\theta<1/\Gamma_X\sim0.1$, a large scatter would be introduced to the $E_\gamma/E_{\rm fb}$ ratio. The absence of such large scatter implies that the jet's $\gamma$-ray emission does not vary significantly over $0.01<\Delta\theta<0.1$ angular scales within the jet opening angle $\theta_j$.

\subsection{Variability in the microscopic parameters of the afterglow model: $\varepsilon_e$, $p$} \label{sec:micro}
Significant burst to burst variations in $\varepsilon_e$ and/or $p$ would lead to a significant scatter in the $E_\gamma/E_{\rm fb}$ ratio, while significant dependence of $\varepsilon_e$ and/or $p$ on the shock Lorentz factor would lead to a significant deviation from unity of the ratio $E^{11\rm hr}_{\rm fb}/E^{3\rm hr}_{\rm fb}$. Our results therefore indicate that $\varepsilon_e$ and $p$ are uniform between bursts, which is reasonable given the fact that they are determined by the micro-physics, and that their values do not strongly depend on the shock Lorentz factor.

We note that some earlier studies have found significant scatter in $\varepsilon_e$. For example, \cite{y+03} find values in the range of $0.12-0.34$ by fitting multiband observations to 4 different afterglows, and \cite{bkf04} find a factor of 4 difference between the two afterglows for which they present their analysis. While the intrinsic variance we find does not exclude such scatter in $\varepsilon_e$, it would imply that almost all the intrinsic variance we find in the $E_\gamma/E_{\rm fb}$ ratio is due to the variance in $\varepsilon_e$. This, in turn, would imply that the variance due to burst-to-burst variations in $\gamma$-ray production efficiency is negligible. Since $\varepsilon_e$ is a parameter determined by microscopic processes, while the $\gamma$-ray production efficiency depends on the macroscopic properties of the flow (e.g. Lorentz factor variability within the fireball wind, see \S~\ref{sec:efficiency}), this appears to be unlikely.

\subsection{Variability in the GRB $\gamma$-ray production efficiency}\label{sec:efficiency}
Strong burst-to-burst variations in the efficiency of the conversion of the fireball energy to $\gamma$-rays would lead to a large scatter in the $E_\gamma/E_{\rm fb}$ ratio. The absence of such scatter implies that the efficiency is not highly variable between bursts. This is a strong constraint on the fireball model, since significant variations are natural to expect in most of its variants. For example, in models where the emission of $\gamma$-rays follows from the conversion of fireball kinetic energy to internal energy by internal collisions within the wind, the efficiency strongly depends on the amplitude and structure of Lorentz factor variations within the expanding fireball wind. Uniform efficiency suggests an efficient process (with order unity efficiency) of converting fireball energy to radiation.

\subsection{Radiative losses}\label{sec:radiative}
The tight correlation between $E^{11\rm hr}_{\rm fb}$ and $E^{3\rm hr}_{\rm fb}$, and the fact that the ratio $E^{11\rm hr}_{\rm fb}/E^{3\rm hr}_{\rm fb}$ is consistent with unity, implies that no significant radiative cooling occurs between 3 and 11 hrs. In addition, the small intrinsic variance in $E_\gamma/E_{\rm fb}$ implies that if radiative losses are significant at earlier times, they are nearly uniform among all GRBs. These observations can be used to impose constrains on parameters of the fireball afterglow model. We demonstrate this by using the model of \citet{bkf04} for radiative losses, in which the fireball energy drops with time as $E_{\rm fb}\propto t^{-17\varepsilon/12}$ with $\varepsilon=\varepsilon_e/(1+1.05\varepsilon_e)$, as long as all electrons are in the fast cooling regime.

We first note that radiative losses cannot account for most of the scatter between fireball energy estimates at 3 and 11 hours, since the average and std of $\log_{10}(E_{\rm fb, 3\rm hr}/E_{\rm fb, 11\rm hr})$ are $0.04$ and $0.28$ respectively, while radiative losses can only make this ratio larger than unity. Thus, we assume that only the average value is affected by radiative losses, i.e. on average no more than $20\%$ of the fireball energy is radiated away between 3 and 11 hours. This implies either that $\varepsilon_e$ is small, $\varepsilon_e<0.15$, in which case radiative losses are small altogether, or that the stage at which all electrons are in the fast cooling regime ends at $t<1$~hr hour. Using the estimate of \citet{spn98} for the duration of this fast cooling stage, $t_{\rm fast}$, together with our estimate of $\varepsilon_eE_{\rm fb}$, we find $n=2\times 10^{-2}(\varepsilon_B/0.1)^{-2}(\varepsilon_e/0.3)^{-1}(t_{\rm fast}/{\rm 1~hr}){\rm cm^{-3}}$.

To conclude, the small difference ($<20\%$) between the estimated fireball energies at 3 and 11 hours implies either $\varepsilon_e\lesssim0.15$, or a circum burst density $n\lesssim2\times 10^{-2}(\varepsilon_B/0.1)^{-2}{\rm cm^{-3}}$.

\subsection{Energy injection}\label{sec:injection}
The tight correlation between $E^{11\rm hr}_{\rm fb}$ and $E^{3\rm hr}_{\rm fb}$, and the fact that the ratio $E^{11\rm hr}_{\rm fb}/E^{3\rm hr}_{\rm fb}$ is consistent with unity, implies that no significant injection of energy takes place during this time period. Can significant energy injection at yet earlier time be a common feature of GRBs?

Since some the afterglow lightcurves contain periods of shallower decay than expected by the basic fireball model, it has been suggested that additional energy might be injected to the fireball at times later than the initial burst. Based on their analysis of 9 lightcurves (out of a sample of 27) containing a phase of shallow decay, \cite{n+06} suggest that energy injection might increase the fireball energy by a factor $\ge4$ on a timescale mostly up to $\approx3$ hours. In fact, their analysis (see their table 3) suggests that for these burst, the increase in fireball energy can even be much more significant, on the order of a factor of 10 on average (which, if taking into account radiative losses which occur simultaneously, could actually correspond to an even more significant energy injection). The variance in the factor by which the energy is increased is $\sim2-6$.
Similarly, \cite{z+06} present lightcurves containing a shallow decay phase, and interpret them as indications for late time energy injection, but they do not quantify how representative of the "average GRB" these examples are. In a more recent work focusing on energy injection, \cite{pv12} find that among a sample of $\approx100$ swift afterglows featuring a break in the X-ray lightcurve, $\approx30\%$ of the breaks can be explained as being jet breaks in an adiabatic model, while $\approx60\%$ can be explained as jet breaks in a model including extended energy injection, thus suggesting that energy injection on a time scale of up to $\approx1$ day is a ubiquitous feature of GRB afterglows.

Our results, on the other hands, favor the possibility that late time energy injection is not a significant feature in GRBs. First, the relatively low intrinsic variance which we find in the ratio of the prompt energy and of the fireball energy, $E_\gamma/E_{\rm fb}$, implies that any significant energy injection must be correlated with the prompt emission, i.e. it should increase the fireball energy by a constant factor for all bursts. While we cannot rule out such a possibility, it would mean that the already challenging high efficiency of conversion of fireball energy to $\gamma$-rays should be even higher, which would then be difficult to explain. Moreover, \cite{pv12} find a large scatter, of $\approx2$ orders of magnitude, in the factor by which energy injection increases the fireball energy,
which is inconsistent with the small variance we find in $E_\gamma/E_{\rm fb}$.

Second, the significant energy injection inferred by \cite{pv12} during the time period of 100~s to 100~kilo-s implies an average increase in energy by a factor of $1.5-3.5$ between 3 hours and 11 hours, inconsistent with our finding that $E^{11\rm hr}_{\rm fb}$ and $E^{3\rm hr}_{\rm fb}$ are the same on average. Third, their analysis is based on detecting breaks in the lightcurves, which are interpreted as jet breaks which occur at $t<11$~hr for 60\% of the cases. This is inconsistent with our finding that jet breaks do not typically occur at such early time.

\begin{figure}
\epsscale{1} \plotone{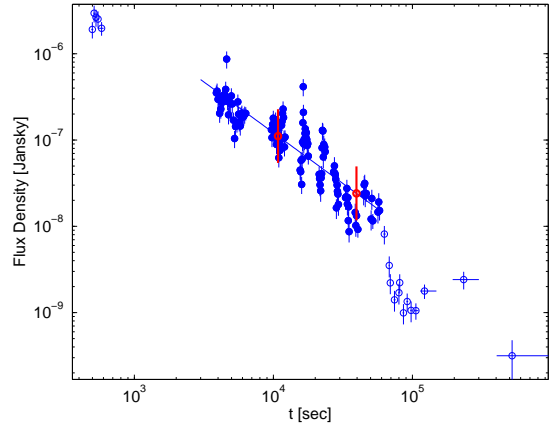}
	\caption{XRT data for GRB090516. The straight line is the power law fit, from which the flux values at 3 and 11 hours are inferred (red circles). The uncertainties in the inferred values (red lines) are based on the scatter of the X-ray data points. Open circles denote data points at times which are outside the time window which we use for determining the X-ray fluxes at 3~hr and 11~hr ($3\times10^3$~s to $6\times10^4$~s for all bursts in the sample).}
	\label{fig:xrt_090516}
\end{figure}

Finally, looking specifically at the bursts common to our sample and to the sample of \cite{pv12}, and which feature a $\ge3$ hours break, which according to their analysis can only be explained with the presence of late time energy injection, we find the estimated fireball energies at 3 and 11 hours to be similar. As an example, the XRT data for GRB090516, for which \cite{pv12} find a break at $2\times10^4$~s, is shown in figure~\ref{fig:xrt_090516}. In our analysis, the energy estimates at 3 and 11 hours are similar for this GRB, and the break, which is not analysed in the current work, appears to occur later.

\subsection{Inverse-Compton losses}\label{sec:IC}
If the electrons responsible for the emission of X-rays via synchrotron emission lose a significant fraction of their energy by inverse-Compton (IC) emission at much higher frequencies, this would lead to a significant under estimate of the fireball energy $E_{\rm fb}$. As was already noted by \citet{bkf03}, since the effect of IC losses depends strongly on $n$ (and on $\varepsilon_B$), we would expect it to vary strongly between bursts, leading to a large scatter in $E_\gamma/E_{\rm fb}$. The small scatter in $E_\gamma/E_{\rm fb}$ suggests that IC losses are not significant at 3 and 11~hrs. We do not carry a detailed analysis of the implications of this conclusion to fireball model parameters, but note that IC emission may be suppressed by the Klein-Nishina effect for reasonable model parameters~\citep[e.g.][]{nas09}.

\section{Conclusions}\label{sec:conclusions}

We have analyzed a sample of 91 \textit{swift} Gamma-Ray Bursts (GRBs) with known redshifts, and found $E_\gamma$ and $E_{\rm fb}$ to be tightly correlated and $E^{3\rm hr}_{\rm fb}$ and $E^{11\rm hr}_{\rm fb}$ to be tightly correlated (\S~\ref{sec:results}). The average(std) of $\eta^{11\rm hr}_\gamma\equiv\log_{10}(E_{\gamma}/(3\varepsilon_eE^{11\rm hr}_{\rm fb}))$ are $-0.34(0.60)$, and the upper limit on the intrinsic spread of $\eta_\gamma$ is approximately 0.5. If the uncertainties in the determinations of $E_\gamma$ and $E_{\rm fb}$ are twice larger than we estimated, then the data imply no intrinsic variance in $\eta_\gamma$. The average(std) of $\log_{10}(E^{3\rm hr}_{\rm fb}/E^{11\rm hr}_{\rm fb})$ are $0.04(0.28)$. Given our estimated uncertainties in inferring $E_{\rm fb}$, the data are consistent with no intrinsic variance in the ratio $E^{3\rm hr}_{\rm fb}/E^{11\rm hr}_{\rm fb}$.

The implications of these results were discussed in detail in \S~\ref{sec:discussion}. The small variance of $\eta_\gamma$ implies that burst-to-burst variations in $\varepsilon_e$ and in the efficiency of fireball energy conversion to $\gamma$-rays are small, and suggests that both are of order unity. It also implies that burst-to-burst variations in $p$ are small. The small variance of $\eta_\gamma$ and the similarity of $E^{3\rm hr}_{\rm fb}$ and $E^{11\rm hr}_{\rm fb}$ further imply that $\varepsilon_e$ and $p$ do not vary significantly with shock Lorentz factor, and that for most bursts the modification of fireball energy during the afterglow phase, by processes such as radiative losses or extended duration energy injection, are not significant (except possibly for a minority, $\sim5\%$, of the bursts). Finally, our results imply that if fireballs are indeed jets, then the jet opening angle satisfies $\theta\geq 0.1$ for most cases.

\textbf{\acknowledgments}
This work was partially supported by the I-CORE Program of the UPBC (grant No. 1937/12) and by a Pazi foundation grant. The work of MAM was partially supported by the Kupcinet-Getz Program at the Weizmann Institute of Science.

\end{document}